# Submission of manuscript to Energy and Buildings

# A numerical approach to evaluating what percentage of a living space is well-ventilated, for the assessment of thermal comfort

*Alain BASTIDE, Philippe LAURET, François GARDE and Harry BOYER*

Contents:

- *Manuscript*
- *List of figures*
- *List of tables*

Corresponding author:

**Alain BASTIDE**

Université de La Réunion

Laboratoire de Génie Industriel, Equipe Génie Civil et Thermique de l'habitat

15 avenue René Cassin, BP 7151, 97705 Saint-Denis Cedex, Ile de la Réunion, FRANCE

tél : 02 62 96 28 90

fax : 02 62 96 28 99

email : alain.bastide@univ-reunion.fr

# A numerical approach to evaluating what percentage of a living space is well-ventilated, for the assessment of thermal comfort

Alain BASTIDE[1], Philippe LAURET, François GARDE and Harry BOYER

*Université de La Réunion, Laboratoire de Génie Industriel, Equipe Génie Civil et Thermique de l'Habitat*

*15 avenue René Cassin, BP 7151, 97705 Saint-Denis Cedex, Ile de la Réunion, FRANCE*

___

**Abstract**

A bioclimatic approach to designing comfortable buildings in hot and humid tropical regions requires, firstly, some preliminary, important work on the building envelope to limit the energy contributions, and secondly, an airflow optimization of the building. For the first step, tools such as nodal or zonal models have been largely implemented. For the second step, the assessment of air velocities, in three dimensions and in a large space, can only be performed through the use of detailed models such as with CFD. This paper deals with the improvement of thermal comfort by ventilating around the occupants. For this purpose, the average velocity coefficient definition is modified to be adapted to CFD and the areas involving movement or the living spaces. We propose a new approach based on the derivation of a new quantity: the well-ventilated percentage of a living space. The well-ventilated percentage of a space allows a time analysis of the aeraulic behaviour of the building in its environment. These percentages can be over a period such as one day, a season or a year. These kinds of results are helpful for an architect to configure the rooms of a house according to their uses, the environment, the architectural choices and the constraints related to the design of bioclimatic buildings.

___

[1] Corresponding author.

E-mail address : alain.bastide@univ-reunion.fr (A. Bastide)

# 1. Introduction

1.1. World context

The energy situation in emerging and insular countries is becoming alarming. The demand for electric power continues to grow whereas the means of production remain limited, and the use of air conditioning (in order to improve thermal comfort) during the hot season exacerbates the problem. A great number of these countries are in the inter-tropical zone and are thus subjected to high temperatures and humidity most of the year. These climates and the increase in the purchasing power of the populations lead to greater use of air-conditioners.

However, the electric power produced from fossil energies such as coal, oil, or gas, or from uranium, will disappear in the coming decades. Before the disappearance of these resources, galloping inflation, due mainly to the scarcity of these fuels, will make their purchase at reasonable prices impossible. It will then become too expensive to operate these air conditioning systems.

The French government [1] and the European Union [2] plan to reduce fourfold their $CO_2$ emissions over the next few decades, and the building sector is one of the principal energy consumers. For this reason, a particular effort is being made so that the buildings in Northern Europe consume low electric power during the cold season. In the tropical Ultra-Peripheral Regions (UPR), this objective of cost reduction is adapted to the local climatic constraints. The reduction of the energy costs related to buildings is thus centred on the reduction of the use of air-conditioning during the year.

To achieve this goal, it is necessary to design comfortable buildings which do not use, or hardly use, active systems. One of the alternatives is to design buildings using the minimum of fossil energy resources in a direct or indirect way, by supporting passive systems such as ventilation [3]. Ventilation offers two advantages. - lower building energy consumption, and an increase in the occupants' thermal comfort [5]. This paper deals with this second effect of ventilation.

1.2. Ventilation

In tropical countries where the air is particularly humid, thermal comfort does not depend solely on cooling rooms but also on the air motion near the occupants. Optimal air velocities for thermal comfort defined in the literature lie between $0.3 m.s^{-1}$ and $0.7 m.s^{-1}$ for moderate activity [6]. Nevertheless, the maximum limit is higher for more intensive physical activities. In the case of office activities, as the sheets of paper on a desk start flying around at about 1m/s, the optimal (and more restrictive) range for low activity lies between $0.3 m.s^{-1}$ and $0.7 m.s^{-1}$ [7, 8].

1.3. Tools for ventilation evaluation

Boundary layer wind tunnel or numerical wind tunnel experiments have been undertaken by many authors. These authors focused their studies on measurements or evaluations of velocity at fixed heights (close to 1.5m [11, 10, 9, 6] and 1.0m [6]) to conclude on the performance of openings. However, Prianto *et al* [6] observed variations of the $C_V$ coefficient depending on two defined heights. This led us to analyze velocity variations as a function of height, and then to particularize our study with sub-volumes of the room volume. These sub-volumes are the living spaces. To estimate ventilation in these spaces, we introduce a new quantity: the well-ventilated percentage of living space. This quantity is inferred from the results of CFD simulations. In the following, the well-ventilated percentage of living spaces is used to estimate ventilation in twelve living spaces. It is also used to compare the performance of various opening distributions.

1.4. A numerical approach

This study involves an adapted experimental protocol. The *in situ* experimentation is very useful for analyzing and understanding the airflows in buildings. Nevertheless, this kind of experimentation can be expensive because of the prices of probes, data-loggers and buildings for observing velocities in three dimensions. The weather data are not controlled and it is difficult to link these data to the indoor air distribution. Moreover, the number of case-studies is limited.

Computational Fluid Dynamics (CFD) is well-adapted to observing the flow pattern inside buildings in a controlled environment. This pattern can be investigated and treated to produce quantitative information on the ventilation of buildings. The advantage of this experimental approach is one of cost: it is less expensive than *in situ* experimentation and can treat a large range of buildings, environments and weather conditions. The principal difficulty is choosing a CFD model which computes accurate inside and outside velocity fields with a computing time which is reasonable for engineers and architects. In this paper, a RANS model is applied to obtain the velocity fields. The methodology presented in the following requires large simulation time and disk space, which need to be reduced. An intelligent coupling strategy is proposed to reduce the computing time.

**2. Numerical methods**

2.1. Field of study

Fig. 1: Sketch of the test building in section (left) and plan (right)

The building tested (Figure 1) is a cubic building of side $3.2m$. A first set of openings is defined. They are squares and are located at the centres of the external frontages. The thickness of the walls is $0.1m$. The thickness of the ceiling is $0.3m$. The building is at the centre of a rectangular field of dimensions $30m$ by $20m$. The volume of the room is given by a square base of side $1.5m$ and a height of $2.8m$.

2.2. Discretization

A grid based on the given dimensions was set up. This grid is coarse far from the walls, but near to the walls and inside the building the grid is refined. Three grids per type of building made it possible to define an optimum grid in a number of cells with respect to the quantities observed for this study.

2.3. CFD modelling

Several assumptions are required when using CFD. Firstly, buoyancy effects are neglected. This assumption was used by Kindangen [11], Gouin [10], Sangkertadi [12] and Ernest [9] in their studies on building ventilation optimization in humid tropical environments. Secondly, any surrounding ground is considered unobstructed, and lastly, it is necessary to choose a turbulence model which is adapted to the resolution of the turbulent field [13]. For this purpose, RNG-k-ε was used [14]. The atmospheric boundary layer is modelled according to a logarithmic law. The ground roughness is $0.077m$, which corresponds to an unobstructed plane [9]. The indoor and external walls of the building are considered smooth.

**3. Classical methods of ventilation evaluation**

3.1. Coefficient of velocity correlation

A great number of results have been obtained by using boundary layer wind tunnel experiments [9, 10]. Others, more recently, were found using numerical fluid mechanics [6, 11]. All these experiments were performed according to the same experimental protocol. Starting from a reference building, velocity measurements of air motion are carried out on a horizontal plane and at a fixed altitude ($z_{ref}$) defined by the experimenter. The ventilation optimization is obtained by modifying the reference building. These modifications generally take place on the building envelope. The quantity used to observe the improvements in ventilation of the building is the average velocity coefficient $C_V$ (2):

$$C_{V,i} = \frac{U_i}{U_{ref}(z_{ref})} \quad (1)$$

$$C_V = \frac{1}{N}\sum_{i}^{N} C_{V,i} \quad (2)$$

The coefficient $C_V$ is the average of the coefficients $C_{V,i}$ (1) evaluated for each point of measurement. Further, the coefficient $C_V$ was linked to many geometrical and weather parameters. The geometrical parameters are the frontage porosity [9], the roof shape [11], the number of openings, the building shape [9, 11], and the height of the ceiling [11]. The weather parameters are the incident angle of the wind compared to a reference axis and the reference velocity $U_{ref}(z_{ref})$.

Fig. 2: Test building: representation of the incident angle of the wind and the plane of sensors (dashed square) and position of probes (points)

For measurements at a fixed height (Figure 2), the average velocity is calculated using specific velocity measurements. This method possesses two main drawbacks. Firstly, it requires a great number of probes in order to have a representative value for the average velocity in a horizontal plane. Secondly, the average velocity can only be measured for a fixed height ($z_{ref}$). In other words, this method does not take into account the velocity variations as a function of height.

3.2. Velocity profile according to height

Fig. 3: Profile of average velocity coefficient according to height inside a test building.

As an illustration, a profile of the average velocity coefficient calculated for a set of horizontal planes is shown in figure 3. The horizontal dashed lines represent the lower and higher opening limits. The vertical dotted lines represent the lower and higher domain limits of $C_V$.

Figure 3 shows that the average velocity coefficient is strongly height-dependent. In the present case (the buildings of figure 1 and 2), the maximum value of the average velocity coefficient is located in a plane passing through the openings, at a height close to 1.1m. It decreases gradually to zero on the levels of the ground and the ceiling. Between the lower and higher limits of the openings the value of the coefficient $C_V$ exhibits variations of almost 160%. In general, variations of over 200% have been observed.

**4. A new adapted model**

The previous section highlights the limits of the existing $C_V$ models. In addition, a specific experimental set-up may be expensive. In our opinion, one has to rely on numerical experiments in order to overcome these limitations. Thus, in order to improve the $C_V$ models, we propose in this paper a new approach that consists in evaluating ventilation in a portion of the room. Furthermore, it will be shown that the method is adapted to the occupant's areas of movement.

4.1. Areas involving movement

Areas involving movement are defined by several parameters related to their activity and to the number of occupants included in the volume of study. For example, in a classroom where the students sit at their tables, the movement area lies between the heights of 0m and 1.5m. These distances can be adjusted by taking into account the fact that certain parts of the body are more sensitive than others to air motion. We therefore consider that the parts of the body where ventilation must be optimized are the occupants' upper bodies and heads. Further, the occupants of this room are at a distance of at least 0.3m from the vertical walls. A living space is defined to take into account all these parameters in the study of a naturally ventilated building in a humid tropical climate.

4.2. Average velocity coefficient

The average velocity coefficient $C_V$ given by equation (1) must be adapted to the grid of calculation. Indeed, the resolution of the velocity field by the RNG-k-ε model equations makes it possible to know the average velocity in each cell. Since the grid is adapted to the geometry of the physical problem, the grid is irregular or unstructured. It is then necessary to take into account the mean velocities weighed by volume.

$$\widetilde{C}_{V,i} = C_{V,i} \frac{v_i}{v_{tot}} \qquad (3)$$

$$\widetilde{C}_V = \sum_i \widetilde{C}_{V,i} \qquad (4)$$

The equation (4) for the average velocity coefficient takes account of the volume ($v_i$) of the cell $i$ where the calculated average velocity coefficient is $C_{V,i}$. The total volume ($v_{tot}$) then represents the sum of all the cells included in the living space.

The coefficient $C_{V,i}$ is generally assessed from velocities of a fixed and single reference ($z_{ref}$): the roof height [11], or the evaluation plane height [6]. However, to compare the ventilation effectiveness at various heights in the building, for several envelope configurations, and for the evaluation of the average velocity coefficient in a living space, an external height-independent reference for the building must be defined.

4.3. Height-independent reference velocity

The reference velocity comes from the atmospheric boundary layer velocity profile. We used a logarithmic profile curve given in equation (5):

$$U(z) = \frac{u_\tau}{\kappa} \log\left(\frac{z}{z_0}\right) \qquad (5)$$

The velocity $U(z)$ is measured for height z = 10m on a site with roughness $z_0$. The values of $U(z)$, $z$ and $z_0$ for the studied site are known, and the $u_\tau / \kappa$ value can thus be calculated. The $U(z)$ expression is then:

$$U(z) = \frac{U(10)}{\log\left(\frac{10}{z_0}\right)} \log\left(\frac{z}{z_0}\right) \qquad (6)$$

The selected reference velocity is then $U_{N,ref}$ defined below:

$$U_{N,ref} = \frac{U(10)}{\log\left(\frac{10}{z_0}\right)} \qquad (7)$$

4.4. Modified and adapted average velocity coefficient

The coefficient $\widetilde{C}_V$ is then given by the following expressions:

$$\widetilde{C}_{V,i} = \frac{U_i}{U_{N,ref}} \qquad (8)$$

$$\widetilde{C}_V = \sum_i \widetilde{C}_{V,i} \frac{v_i}{v_{tot}} \qquad (9)$$

This new coefficient $\widetilde{C}_V$ then allows the evaluation of the mean velocity in a building from measured weather data, in our case the airflow at a height of 10 meters. Nonetheless, the disadvantage of this kind of parameter is that it does not provide information on the velocity distribution inside a living space. It must be noted however that Ernest *et al* [9], Gouin [10] and Kindangen *et al* [11] made use of the standard deviation provided by the empirical distribution.

However, the mean velocities as well as the standard deviations depend on the occupation of the dwelling. As a consequence, the search for the best envelope configuration can prove difficult.

Thus, we propose a single output model that combines optimum velocities, living spaces and weather data. The single output of this model represented by a percentage must give information about the air velocity distribution in the living space and must provide easily exploitable non-dimensional data.

4.5. Well-ventilated percentage of space

Our methodology is based on the study of the percentage of the volume in a living space where the air velocity evaluated from the CFD is acceptable to provide thermal comfort.

In a living space, ventilation is suitable if the air velocity is included in a velocity *range*. The boundaries of this range are noted $U_{min}$ and $U_{max}$. The percentage P of well-ventilated space is therefore the volume in which the velocity of each cell lies between these two values ($U_{min}$ and $U_{max}$) divided by the total volume ($v_{tot}$) of the living space. The equation then takes the following form:

$$P(U_{min} < U_i < U_{max}) = \sum_{i}^{U_{min} < \widetilde{C}_{V,i}.U_{N,ref} < U_{max}} \frac{v_i}{v_{tot}} \qquad (10)$$

The velocities $U_i = \widetilde{C}_{V,i}.U_{N,ref}$ (Equation 10) are evaluated for the data of a selected weather sequence. This percentage $P$ takes into account the evaluated reference velocity starting from the weather data and living space defined by the modeller.

The computation of P requires a great number of simulations and a large amount of data. The method is based on the flow characteristics and non-dimensional quantities; a specific coupling strategy of the CFD results with the weather data dramatically reduces the number of CFD simulations.

4.6. Coupling strategy

Not all the CFD data are stored in a data base (Figure 4). We used the fact that with high Reynolds number ($Re > 40000$), the form of the flow is Reynolds number-independent [15, 16] so that the velocities evaluated in each cell are proportional to the reference velocity $U_{N,ref}$.

Consequently, the incident angle is the sole weather parameter to categorise the data. The space co-ordinates which define a living space are used to select the cells in the CFD database which are relevant for the calculation of the well-ventilated percentage of space. Once the CFD cells have been selected, the velocity in each cell $U_i$ is modified according to the reference velocity $U_{N,ref}$ (see Equation 7). The calculation of the percentage is then performed.

Fig. 4: Data organization

**5. Variations of the adapted average velocity coefficient in various living spaces**

The choice of a living space is conditioned by the *studied* level of occupation of the room, the activity of the occupants, their sizes and the ventilation of the part of the body which the modeller wishes to optimize. The influences of furniture and people are supposed to be negligible [9-11].

5.1. Definition of various adapted volumes

Table 1 gives a certain number of defined domains inside the building.

Table 1: Movement areas defined in the volume of the room

Each domain or living space is numbered; the coordinates of living spaces are given according to the positions of the extreme points $(X,Y,Z)$. Lastly, a plan view (H) and section (V) of the living space relative to the coordinates is shown.

5.2. Justification of the choice of dimension

The field D1 represents the total volume of the room. To show the influence of the flow close to the walls, a volume (D2) similar to the D1 volume is defined. This latter excludes the parts of volumes where the occupants do not move i.e. close to the walls, ground and ceiling. The ventilation of the person's trunk is studied using volume D4. The ventilation of children is evaluated by the D3 volume. The ventilation of a person confined to bed, and therefore near to the ground, is observed using volume D5. A person, working upright, like a teacher, is studied *starting* from volume D6. This volume D6 is adjusted to observe ventilation only at the level of the trunk using volume D8. The well-ventilated percentage of space, with a small cell height, is comparable to a measurement plane, and is studied starting from volume D7. Lastly, domains D9, D10, D11, and D12 are volumes which highlight ventilation for people sitting at their desks and isolated in a portion of the room.

5.3. Relationships between living spaces and adapted average velocity coefficients

Figure 5 illustrates the effect of including the zones close to the walls or not in the calculation of the adapted average velocity coefficients. The largest $\widetilde{C}_V$ values are for the living space D2.

We find that near to the walls the boundary layers produce lower velocities, and although velocities close to the openings are higher because of the acceleration phenomenon due to the section reduction of the current tubes which go through the building, they do not improve the value of $\widetilde{C}_V$.

Fig. 5: Influence of zones close to the walls (D1 and D2)

The maximum values are obtained in these two living spaces for incident angles close to 20°. The two curves meet for incident angles close to 90°. For these last values, the building configuration is such that the airflow does not directly enter the building. In this case, the evaluated $\widetilde{C}_V$ values result from the air motion due to the turbulent phenomena [9]. The maximum difference between these two cases is observed for angles close to 40°, and the maximum variation between these two curves is then about 29%. In the rest of this document, the cells close to the walls are excluded from the treatment.

In figure 6, the adapted average velocity coefficients in various living spaces are shown to illustrate the influence of height on the living space choice. For domains D7, D4 and D8, the values increase for incident angles going from 90° to 0°. On the other hand, for the domains D6, D3 and D5, the values begin to decrease for respective incident angles from 10°, 20° and 40°. These decreases are due to the form of the flow between the two openings. The incident angles of 40°, 20°, 10°, and 0° are the optimum angles for the ventilation of the corresponding living spaces: D5, D3, D8, D4 and D7.

Fig. 6: Influence of living space height on the adapted average velocity coefficients (Domains 3, 4, 5, 6, 7 and 8)

The flow exhibits its most significant $\widetilde{C}_V$ values in the air volume between the two openings for angles between 0° and 45°. The evaluated air velocity values in the lower part of the room (D3 and D5) are then smaller. The coefficient $\widetilde{C}_V$ is thus smaller in these volumes.

Volume D6 includes both well-ventilated and poorly-ventilated zones. As a result, its $\widetilde{C}_V$ values lie between those of domains D7, D4, D8 and domains D3, D5.

Domain D7 illustrates the traditional method of ventilation evaluation in a plane. The adapted average velocity coefficients in this volume are both more significant than and very different to the representations of the adapted average velocity coefficients in the lower part of the room. Thus, the envelope geometry and the positions of the openings cause most of the ventilation to occur between the openings, to the detriment of the lower part of the room. As can be seen, this method does not allow one to accurately estimate the ventilation in these lower parts of the room.

From 45° to 90°, the air does not flow directly into the building. Air velocity is diffused in the whole room. The amplitude of average velocity coefficients tends to be homogeneous in all the living spaces. The more the angle of incidence tends towards 90°, the more the average velocity coefficients observed tend to be of the same value, and $\widetilde{C}_V$ variations lower than 0.05 are observed. Height therefore has little effect on the choice of living space for angles between 60° and 90°.

Fig. 7: Comparison of the average velocity coefficients for four living spaces (Domains 9, 10, 11 and 12)

Living spaces D9, D10, D11 and D12 are representative of domains containing a seated person. Figure 7 shows various average velocity coefficients according to the incident angle. For four volumes, the maximum coefficient is obtained for incident angles such as when the airflow enters with an angle of ±30° relative to the normal to the wall. These angles are respectively 30°, 150°, 210° and 330° for living spaces D10, D12, D11 and D9. These incident angles allow the air to enter directly into the living space. The minima are observed for angles of 90° and 270°. These angles are such that the directions are parallel to the opening planes, and so the air no longer enters the building directly. Moreover, the building and living spaces possess a plane of symmetry and the results obtained are therefore also symmetric.

Contrary to the preceding case, the null incident angle does not make it possible to observe an optimal value of $\widetilde{C}_V$. On the other hand, for this incident angle, the value of the average velocity coefficients is practically identical in four living spaces. With a null incident angle, we make sure that the ventilation is almost identical in each living space.

5.4. Three different building occupations

From the twelve previous living spaces, three are selected to compare their $\widetilde{C}_V$ values. These three living spaces are related to the use of a room where an occupant can both sleep, work upright and work seated. In figure 8, the variation of these coefficients is represented according to the incident angle. The maximum value is obtained for an incident angle of 210° in the living space D11. The values of the average velocity coefficients in volumes D8 and D9 are then lower by 32% and 54% respectively. Ventilation is thus optimal for an incident angle of 210° in the living space D11. The optimal building orientation therefore for an office positioned in the D11 volume is at 210° compared to the reference (Figure 1).

Fig. 8: Average velocity coefficients ($\widetilde{C}_V$) according to the incident angle for domains D8, D5 and D11.

For incident angles of 0° and 180°, the ventilation is optimal in volumes D8 and D11 whereas the ventilation in the D5 volume is clearly unfavourable. In other words, these two orientations are not favourable to sleeping in this room.

The most favourable orientation for these three living spaces is 210°. Moreover, the best coefficients are observed in living spaces D11 and D5 for a $\widetilde{C}_V$ value of 17% lower than the maximum observed in volume D8.

**6. Variations of the well-ventilated percentages of volumes in various living spaces**

6.1. Definition of a climatic sequence

Figure 9 shows a climatic sequence over one day.

Fig. 9: Hourly evolution of incident angle (left axis) and intensity of velocity reference $U(10)$ (m.s-1) (right axis)

The velocity value $U(10)$ varies between 0.67 m.s$^{-1}$ and 2.16 m.s$^{-1}$. The building is oriented so that the incident angle is 160° during the day and 180° during the night.

6.2. Results: well-ventilated percentages of volumes in living spaces

Figure 10 shows the well-ventilated percentages of volumes in the living spaces given in table 1. These percentages are calculated from the computational fluid dynamics simulation data (see Eq. 10) and from the data related to the external weather conditions shown in figure 9. The three percentages represented are calculated for one period ranging between 8am and 6pm, except for the percentage of volume D5 which is evaluated over one night period (8pm-6am).

Fig. 10: Well-ventilated percentages of volume evaluated during one 24 hour period for 12 different living spaces.

The percentages P1, P2 and P3 represent respectively the percentage of velocities in the ranges 0.3 m.s$^{-1}$ and over, from 0.3 m.s$^{-1}$ to 1.0m.s$^{-1}$, and from 0.3 m.s$^{-1}$ to 0.7 m.s$^{-1}$.

The percentages P1, P2 and P3 are different for eleven domains except for the living space D5. The percentages of P1 are higher than the percentages of P2. This is explained by the fact that the P2 results are included in the results of P1. It is the same for the results for P3 which are included in both P1 and P2. For the given external conditions, we can note that strong variations among the studied volumes are observable for the percentages P1 and P2. The P3 percentages exhibit maximum variations of about 210% between the extremes. These variations are due to the large velocity variations in the room due to the form of the flow. The form of the flow is induced by the position of the openings, the building shape and the incident angle.

6.3. Discussion: three percentages P1, P2, P3

The three percentages are equal in the D5 volume. This means that the evaluated air velocities do not exceed the limit of 0.7 m.s$^{-1}$. On the other hand, for the other living spaces the percentages are different according to their position in the room. The greatest variation between the three percentages is observed in the living space D12. In this living space, the P1 well-ventilated percentage exceeds 90% of the volume whereas the P3 percentage is 47%. So, it is important to note that the results for some living spaces can be very sensitive to the choice of $U_{min}$ and $U_{max}$. In the following, we focus on the most restrictive percentage, the percentage P3.

6.4. Discussion: the ventilation of living spaces

The percentages of living spaces D1 and D2 show the influence of the velocities near the walls or far from the unused domains of the room. The percentages evaluated in volumes D1 and D2 are respectively 55% and 53%. In this case, the influence of the cells close to the walls is weak in this final result.

The airflow enters the room with a direction of 160° during the day and partially crosses living space D12. Although this living space is located at a favourable place for ventilation, it does not benefit fully from its position; it is much less well-adapted than the living space D10, which was the best ventilated living space observed during this simulation. This building is thus well designed for an office-worker. The flow in this living space is relatively homogeneous. The living space D5, however, is the least ventilated - its percentage is 24%. This percentage is evaluated only during the night. Although the reference velocity intensity for this night period is higher than the daytime period, the percentages P1, P2 and P3 are equal. Ventilation in this living space can thus be improved either by modifying the shape and the position of the openings or by erecting the building on a windier site. This building, in this environment, is thus not favourable to the ventilation of a bed-ridden person.

Volume D7 represents the domain presented in the literature. The air cell layer is sufficiently small to be regarded as a plane. We can notice that this layer of cells features a percentage of 49%. It is not representative of the ventilation in the whole room, and in particular of the ventilation in volume D5.

Volumes D9, D10, D11 and D12 were then used to seek the best position for a person working in an office. The best position (D10) features a well-ventilated percentage of volume twice that of the least well-ventilated volume (D11) of the four (D9, D10, D11, D12). An interior designer can thus envisage the optimal position of the door and furniture in the room starting from these results.

The percentages evaluated in living spaces D3 and D5 are very different. The percentages are thus, like the average velocity coefficient, very sensitive to the vertical positions of the living spaces.

**7. Application to ventilation evaluation and optimization**

7.1. Modification of the test building

The ventilation in various living spaces depends on the definition of the domain inside the studied room. We propose to optimise the ventilation in certain living spaces for a weather sequence identical to the preceding one and for different opening geometries.

The buildings and their associated openings are defined in table 2. Buildings 1 and 2 have equivalent frontage porosities with different distributions of openings. The goal of these buildings is to show the advantage of a distribution of openings which has been adapted to the use of the room. The opening positions of building 1 are traditional, whereas building 2 is defined to try and improve the low air velocity distributions in the corner and give better homogeneity of the velocities in the living spaces. Building 3 features a frontage where an opening has been removed. It is intended to observe the influence of Venturi phenomena on the velocity distribution in certain living spaces. Lastly, building 4 makes it possible to simulate building 2 with the large openings closed.

Table 2: Representation of the test building and the opening modifications

7.2. Results: ventilation of a bed-ridden person

Fig. 11: Hourly evolution of the percentage of well-ventilated volume in D5

The time variation of the well-ventilated percentage of volume for the lower part of the room (living space D5) and for each configuration of building is shown in figure 11.

The most favourable technical solution is shape 2. It makes it possible to reach night values ranging between 80% and 92%, for a frontage porosity equivalent to building 1. The most unfavourable solution for the night period (midnight – 7am and 8pm – 11pm) is building 1.

Building 4, which has the smallest frontage porosity, features better ventilation than building 1; the night variation observed reached 160%. At the beginning of the day, configuration 2 remains optimal.

The configurations 1 and 4 display opposite variations to each other through the day. Taking into account the results, configuration 1 is preferable at night whereas configuration 4 is favourable to ventilation during the day.

7.3. Results: ventilation of a person working upright

The living space D8 makes it possible to inform modellers about the ventilation around a teacher standing in front of a table for a whole day, from 8am to 5pm.

Fig. 12: Hourly evolution of the percentage of well-ventilated volume in D8

The best configuration is given by configuration 1. Configurations 2 and 3 are similar whereas building 4 has highly unfavourable ventilation for a person carrying out this activity.

7.4. Results: ventilation of a seated office-worker

In the case of living space (D11), the ventilation is optimised by using the configuration of building 1 during the day. On the other hand, the percentages observed are null in the case of configuration 4, and so the building is particularly badly designed for working in this living space.

The curve relating to configuration 3 shows that the percentage is almost insensitive to the orientation change in the incident angle between 8am and 8pm. Conversely, configuration 2 appears to be sensitive to this orientation change.

Fig. 13: Hourly evolution of the well-ventilated percentage of volumes in D11

## 8. Conclusion

The detailed study of ventilation is thus realizable using the two tools presented in this article: the well-ventilated percentage of volumes and the adapted average velocity coefficient. The models developed on test buildings are very easily adaptable to the study of complex and typical buildings.

Thanks to CFD tools, buildings are easily modified and optimised for *in situ* experimentation. However, an *in situ* experiment is in progress to compare the CFD results with experiments for several weather conditions and buildings.

Living spaces require a coupling strategy of data and CFD results so that calculations are optimal in computing time. Thanks to the strategy detailed in this article and according to the weather data given, the computing time was reduced by a factor of more than ten compared to a traditional approach.

The principal observations and improvements of the models are as follows:

- The velocity variations in the interior of the building are very significant. A study in a measurement plane (i.e. the classical method) is thus limited to concluding on ventilation in a room or in a portion of a room;
- Some living spaces are defined to particularize the study of ventilation to portions of the most used rooms;
- The coefficient $\widetilde{C}_V$ defined in the literature is modified according to the constraints related to the CFD tool. It is in particular adapted to life volumes;
- The well-ventilated percentage of volume is defined and applied to living spaces. It makes full use of all the information connected with the building and its environment to produce a non-dimensional number, easily used by architects and engineers.

The results relating to the coefficients $\widetilde{C}_V$ and the well-ventilated percentage of volume $P$ are:

- The definition of living spaces strongly conditions the results ($P$ and $\widetilde{C}_V$) because of the strong variations in velocity amplitudes in the flow through the building. Future work based on a sensitivity analysis is envisaged;
- The optimization of the room orientation is carried out thanks to the simultaneous study of the $\widetilde{C}_V$ values in several living spaces. An architect can therefore define in advance the optimal arrangement of furniture in a room;

The summary of the results relating to the envelope modifications are as follows:

- The coupling of living spaces and the well-ventilated percentages of volume show clearly that a better distribution of openings improves ventilation in zones initially slightly ventilated.
- We note that the study of the well-ventilated percentage of volume confirms or disproves the results obtained using

coefficients $\widetilde{C}_V$ because they include more parameters related to the building and the environment than the latter.

The aim of future work is the coupling of comfort indices with the thermal building conditions. Indeed, the comfort indices have a range in which the occupants are comfortable. From this optimal range for a thermal comfort index and knowing the hydrous and thermal conditions of a building, an optimal velocity range can thus be obtained. This range is not static (as previously) but dynamic**.** It adapts to the hygro-thermal conditions of the room. The percentage of comfortable volume is then obtained instead of a well-ventilated percentage of volume.

Moreover, if the percentage of comfortable volume is low, because of high air velocities, we are interested in the envelope modifications which could be made by the occupants to reduce the excessive velocity amplitudes. A coupling with adaptive models is then planned to improve the performance of the tools presented in this article. The calculation algorithm must then adapt the building configuration during simulation to take account of these modifications.

**References**

[1] Ministère de l'Ecologie et du Développement Durable - Gouvernement Français. La division par 4 des émissions de dioxyde de carbone en France d'ici 2050, , Rapport de Mission, Ref : Facteur4-VL1, 2004.

[2] Commission des Communautés Européennes. Directive du Parlement Européen et du Conseil sur la performance énergétique des bâtiments, Ref : COM(2001) 226, 2001.

[3] F. Allard, Natural ventilation in buildings: a design handbook, James & James, London, 1998.

[4] Label ECODOM, Operation expérimentale - extension au département de la Guyanne - Prescriptions techniques Document de référence, Cabinet Concept Energie et Promotelec, 1997.

[5] B. Givoni, Man, Climate and Architecture, Elsevier, Amsterdam, Applied Science Publishers Ltd., London, 2nd Edition, 1976.

[6] E. Prianto and P. Depecker, Optimization of architectural design elements in tropical humid region with thermal comfort approach, Energy and Buildings 35 (3) (2003) 273-280

[7] ASHRAE Handbook. ASHRAE Transaction, 2001.

[8] R.M. Aynsley, Effect of airflow on human comfort, Building Sciences 9 (1974) 91-94

[9] D. R. Ernest, F. Bauman and E. Arens, The Prediction of Indoor Air Motion for Occupant Cooling in Naturally Ventilated Buildings, ASHRAE Transactions 97 (1) (1991)  539–552

[10] G. Gouin, Contribution Aérodynamique à l'Etude de la Ventilation Naturelle des Habitats en Climat Tropical Humide, Ph.D. Thesis, University of Nantes, France, 1984.

[11] J. Kindangen, G. Krauss and P. Depecker, Effects of roof shapes on wind-induced air motion inside buildings, Building and Environment 32 (1) (1997) 1-11.

[12] Sangkertadi, Contribution à l'étude du comportement thermo-aéraulique des bâtiments en climat tropical humide - Prise en compte de la ventilation naturelle dans l'évaluation du confort, PhD Thesis, INSA Lyon, France,1998.

[13] A. Bastide, Etude de la ventilation naturelle à l'aide de la mécanique des fluides numérique dans les bâtiments à grandes ouvertures. Application à l'amélioration d'un modèle aéraulique nodal et au confort thermique, Ph.D. Thesis, University of La Réunion, France, 2004.

[14] StarCD, AdapCO, v.3.15, Methodology, London, UK.

[15] A. Bastide, F. Garde, L. Adelard and H. Boyer, Statistical study of indoor velocity distributions for comfort assessment, in: Proceeding of ROOMVENT, Coimbra, Portugal, September, 2004, pp. 1-6.

[16] J.E. Cermak, M. Poreh, J. Peterka and S. Ayad, Wind tunnel investigations of natural ventilation.   J. Transp. Eng. ASCE Trans. 110 1 (1984), pp. 67–79

**Nomenclature**

| | |
|---|---|
| $U_i$ | local velocity magnitude ($m.s^{-1}$) |
| $U_{ref}(z_{ref})$ | reference velocity ($m.s^{-1}$) |
| $U_{N,ref}$ | height-independent reference velocity ($m.s^{-1}$) |
| $C_V$ | average velocity coefficient |
| $C_{V,i}$ | local velocity coefficient |
| $z_{ref}$ | height reference ($m$) |
| $\widetilde{C}_V$ | adapted average velocity coefficient |
| $\widetilde{C}_{V,i}$ | adapted local velocity coefficient |
| $v_i$ | cell volume ($m^3$) |
| $v_{tot}$ | volume of living space ($m^3$) |
| $u_\tau$ | friction velocity ($m.s^{-1}$) |
| $\kappa$ | von Karman's constant |
| $U_{min}$ | min velocity for comfort ($m.s^{-1}$) |
| $U_{max}$ | max velocity for comfort ($m.s^{-1}$) |
| $U(z)$ | velocity profile of atmospheric boundary layer ($m.s^{-1}$) |
| $z$ | height ($m$) |
| $z_0$ | roughness length ($m$) |
| $P$ | well-ventilated percentage of living space |

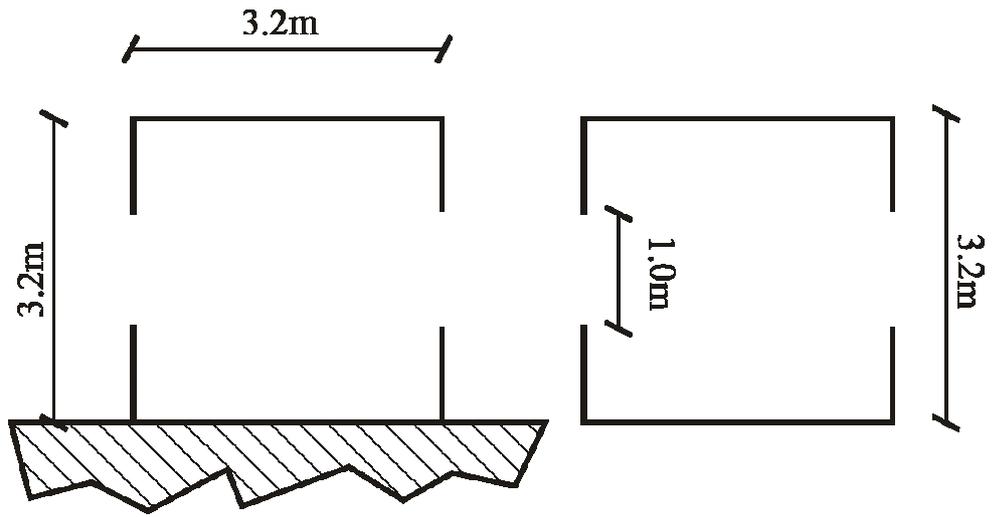

Figure 1: Sketch of the test building in section (left) and plan (right)

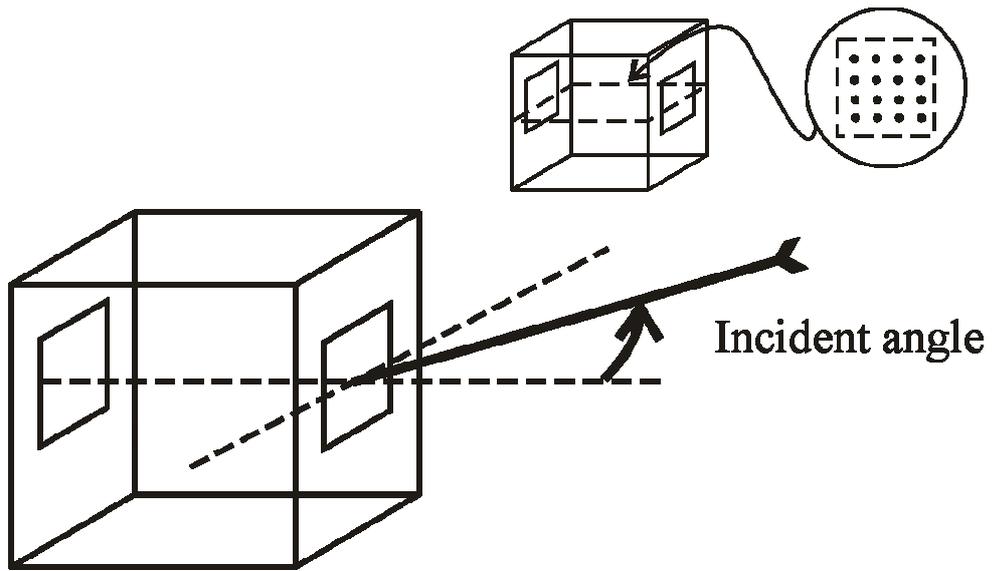

Figure 2: Test building: representation of the incident angle of the wind and the plane of sensors (dashed square) and position of probes (points)

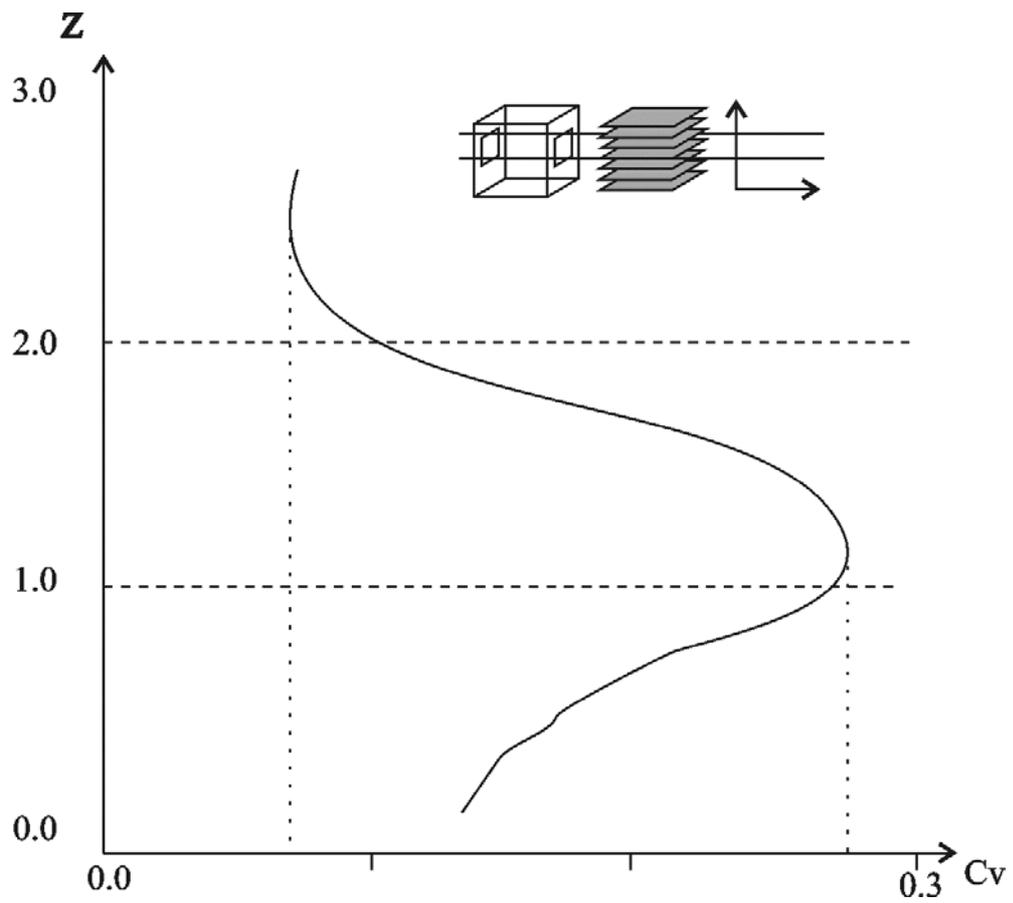

Figure 3: Profile of average velocity coefficient according to height inside a test building.

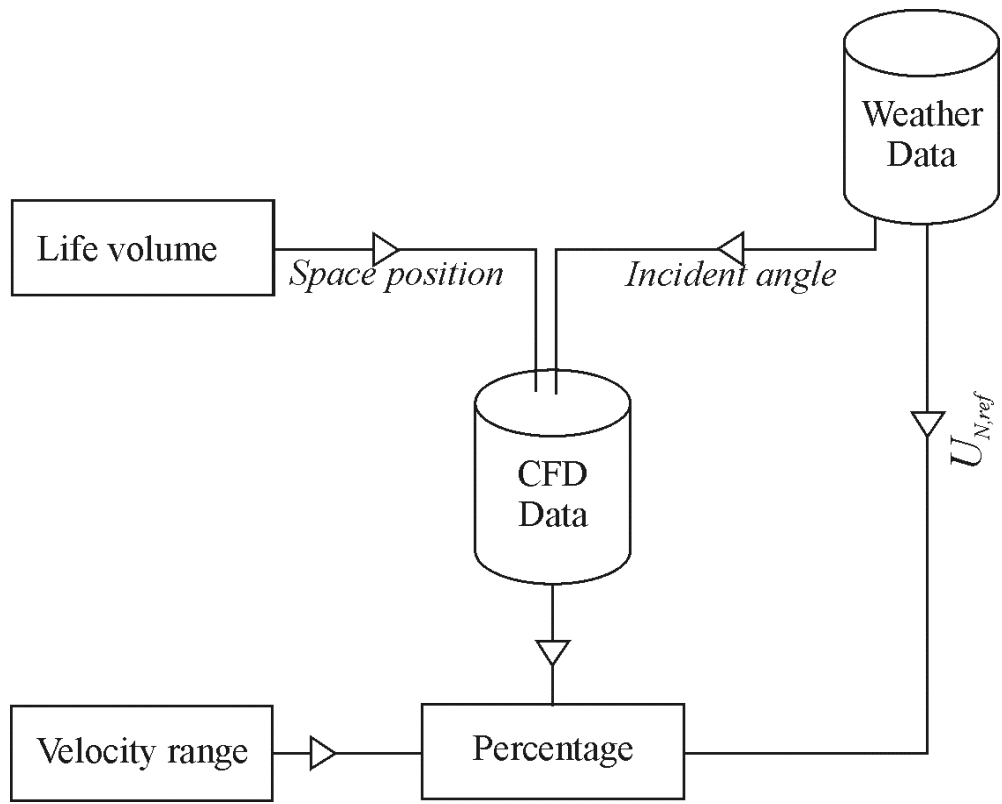

Figure 4: Data organization

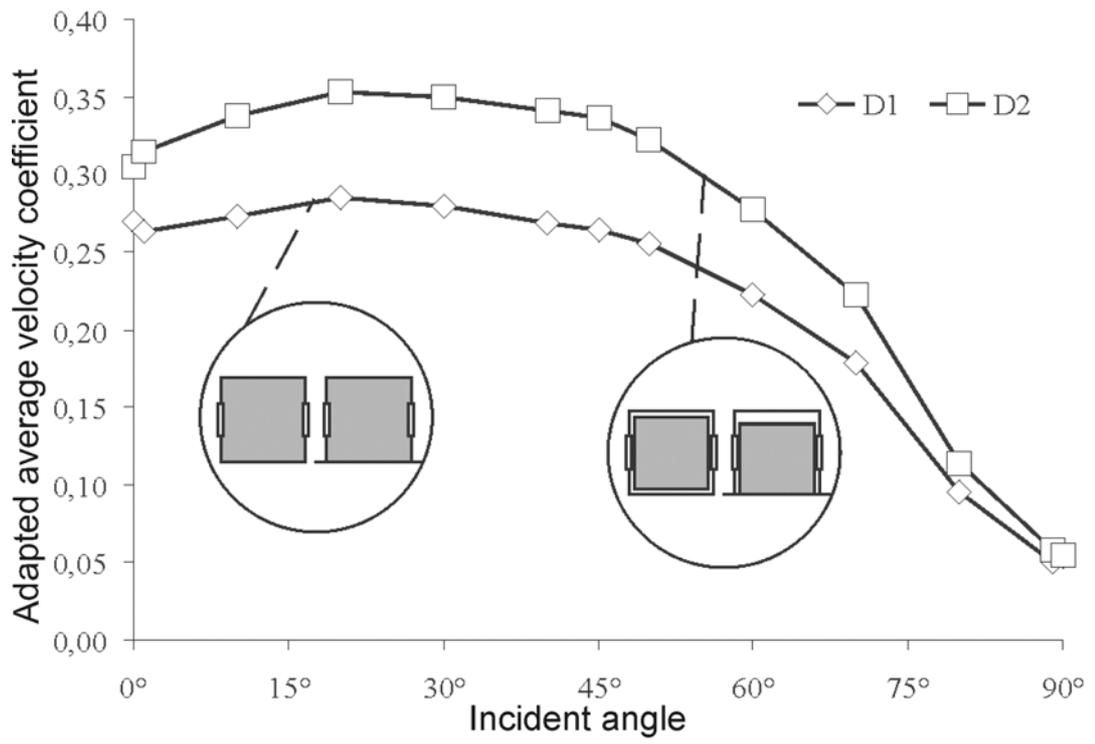

Figure 5: Influence of zones close to the walls (D1 and D2)

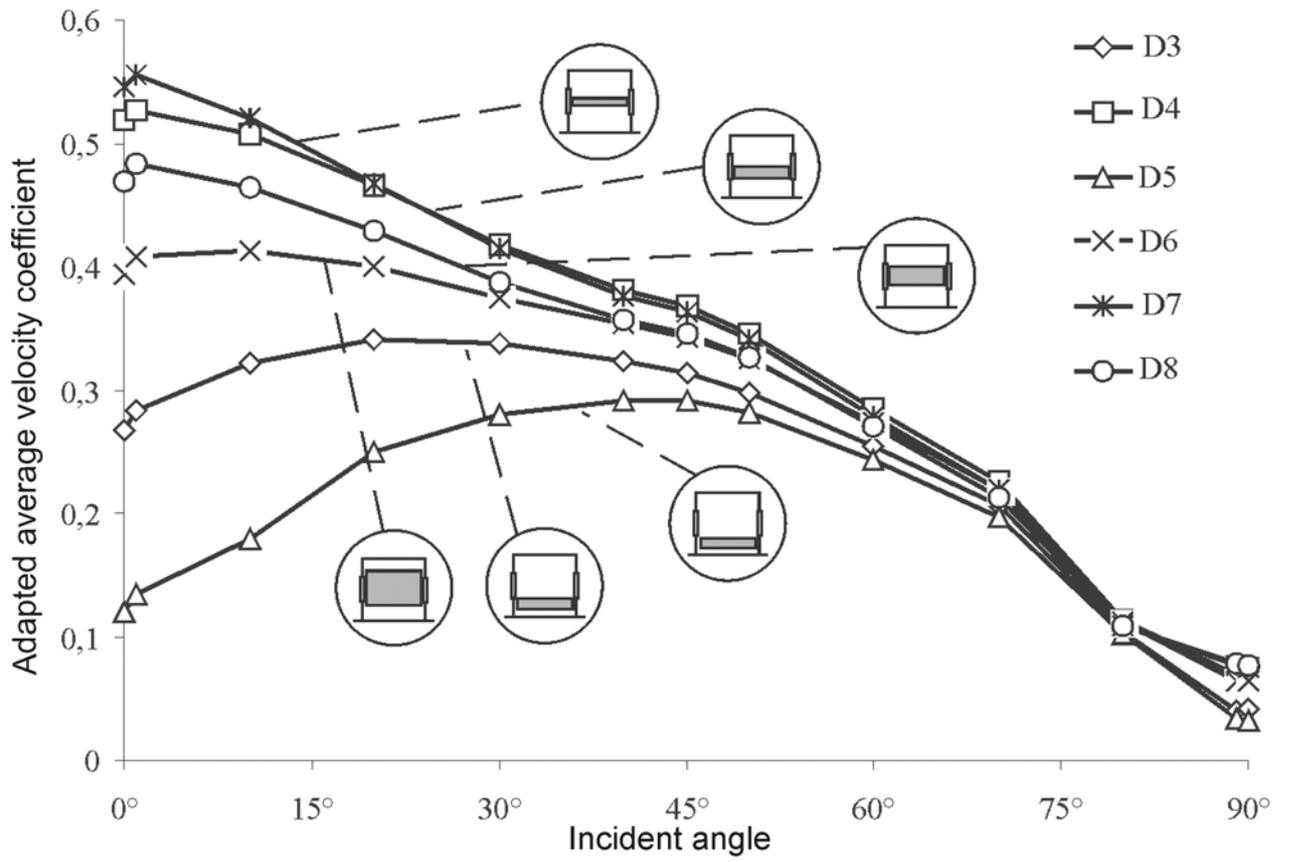

Figure 6: Influence of living space height on the adapted average velocity coefficients (Domains 3, 4, 5, 6, 7 and 8)

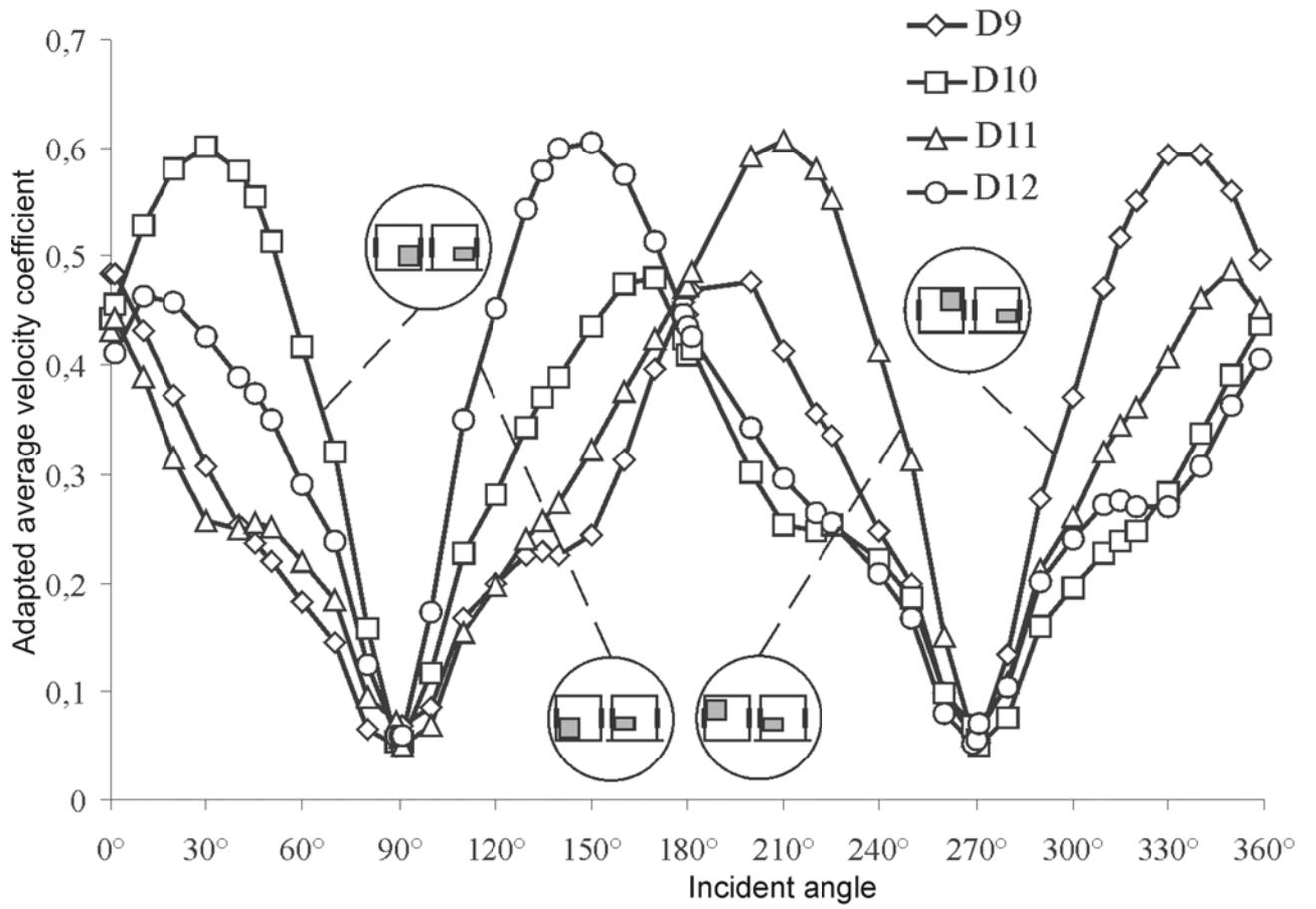

Figure 7: Comparison of the average velocity coefficients for four living spaces (Domains 9, 10, 11 and 12)

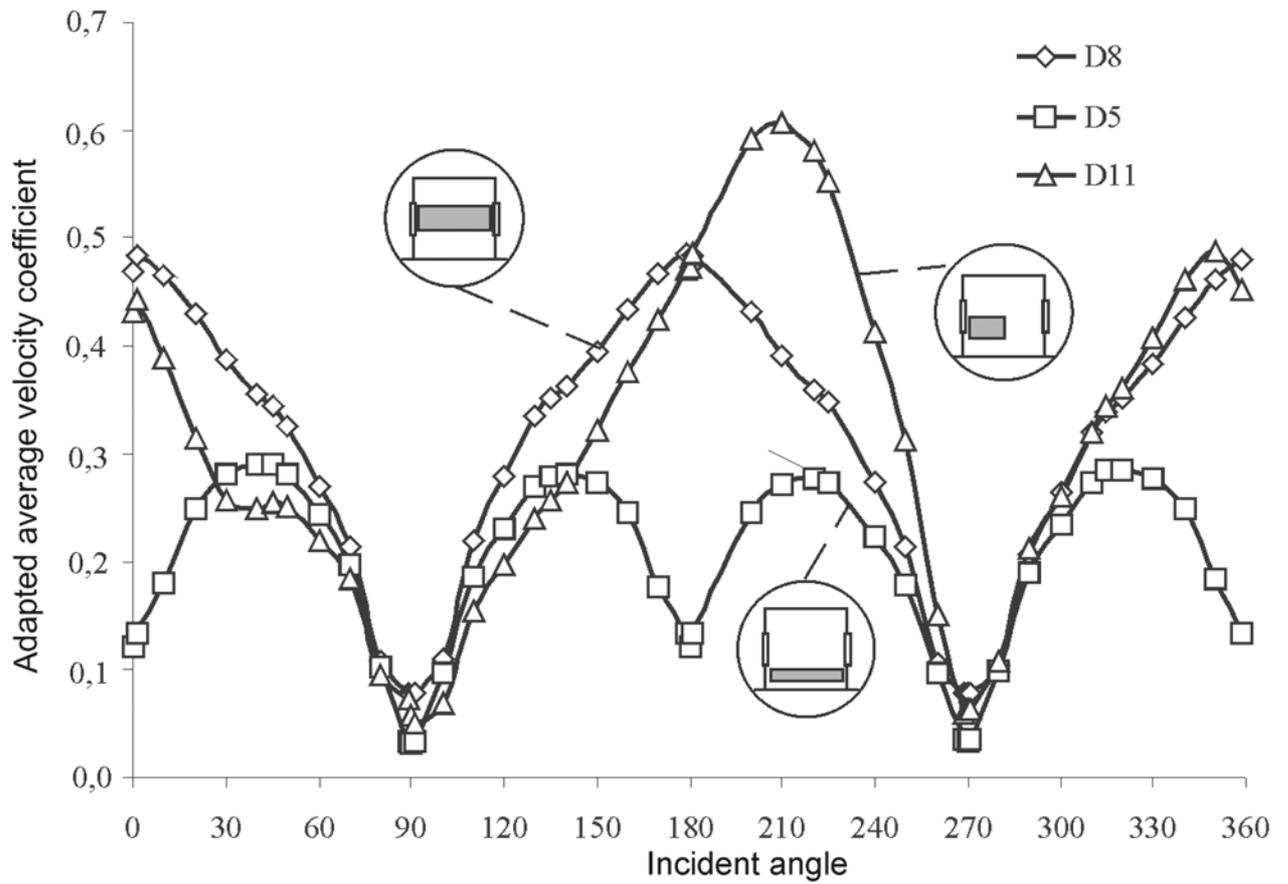

Figure 8: Average velocity coefficients ($\widetilde{C}_V$) according to the incident angle for domains D8, D5 and D11.

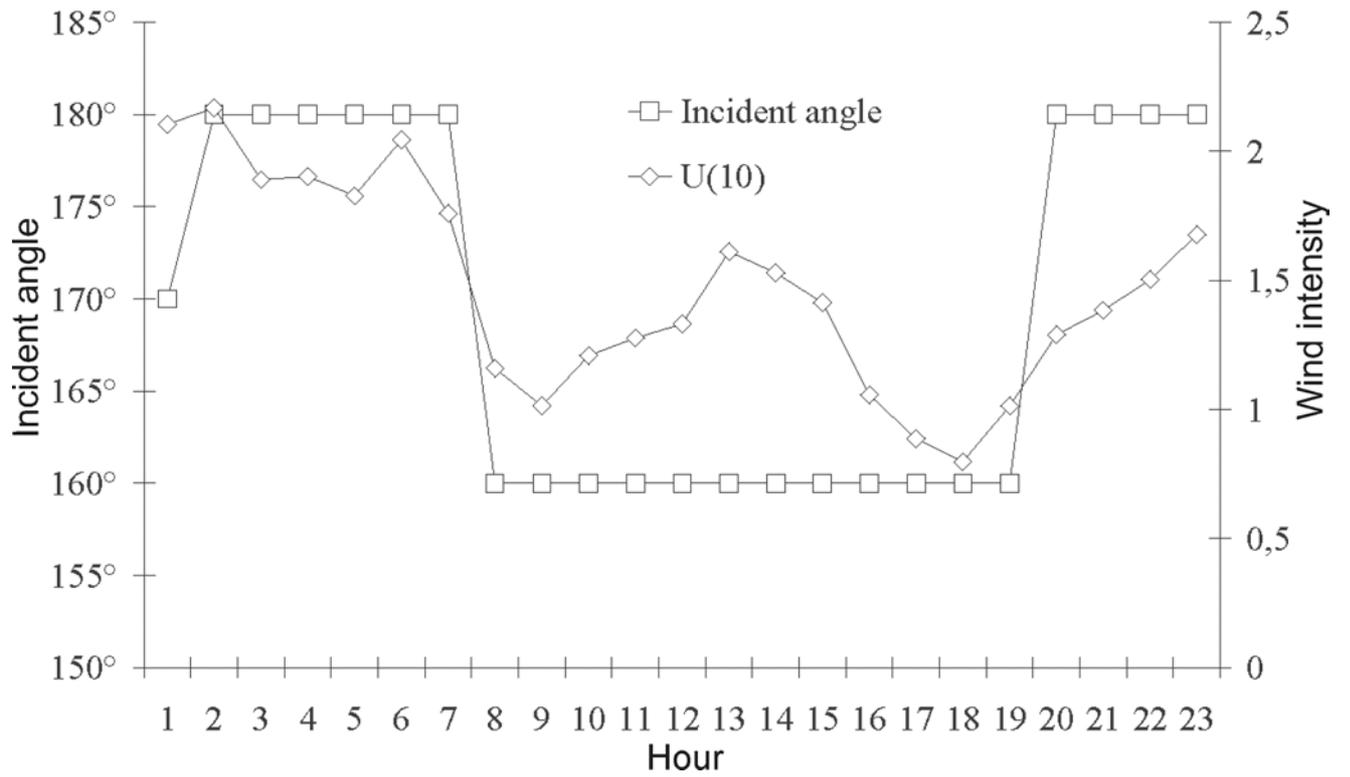

Figure 9: Hourly evolution of incident angle (left axis) and intensity of velocity reference $U(10)$ (m. s-1) (right axis)

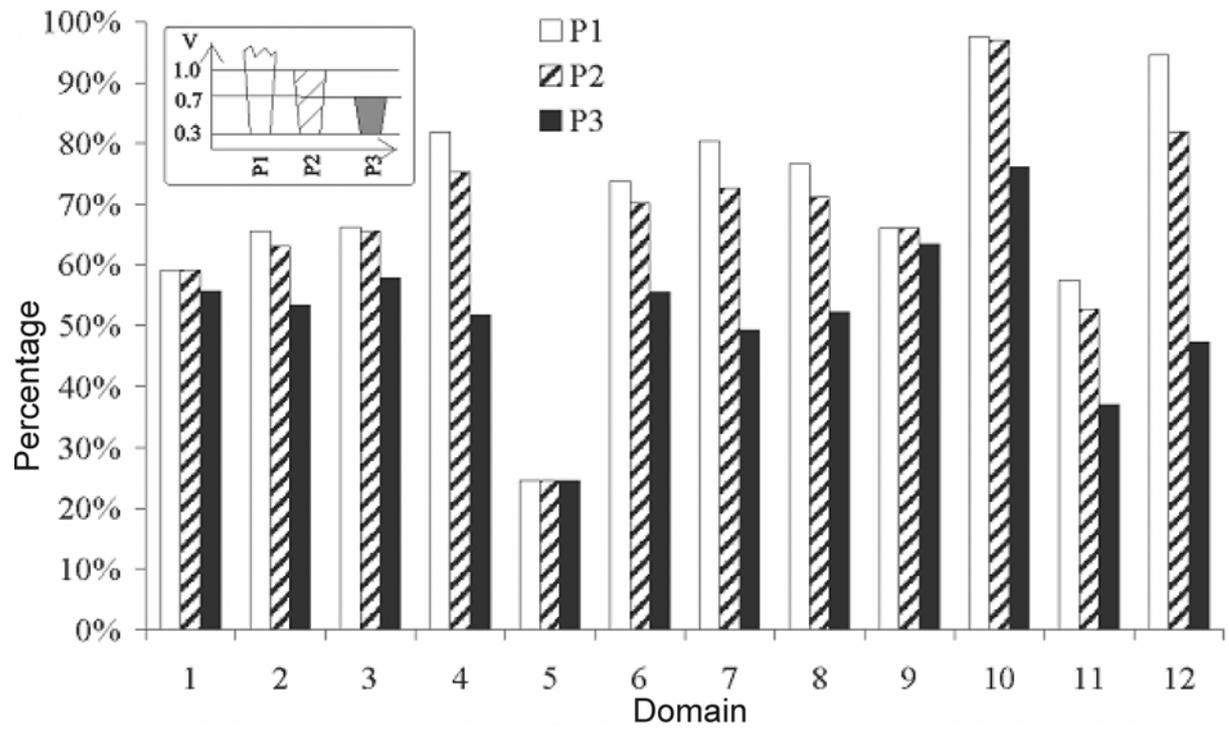

Figure 10: Well-ventilated percentages of volume evaluated during one 24 hour period for 12 different living spaces.

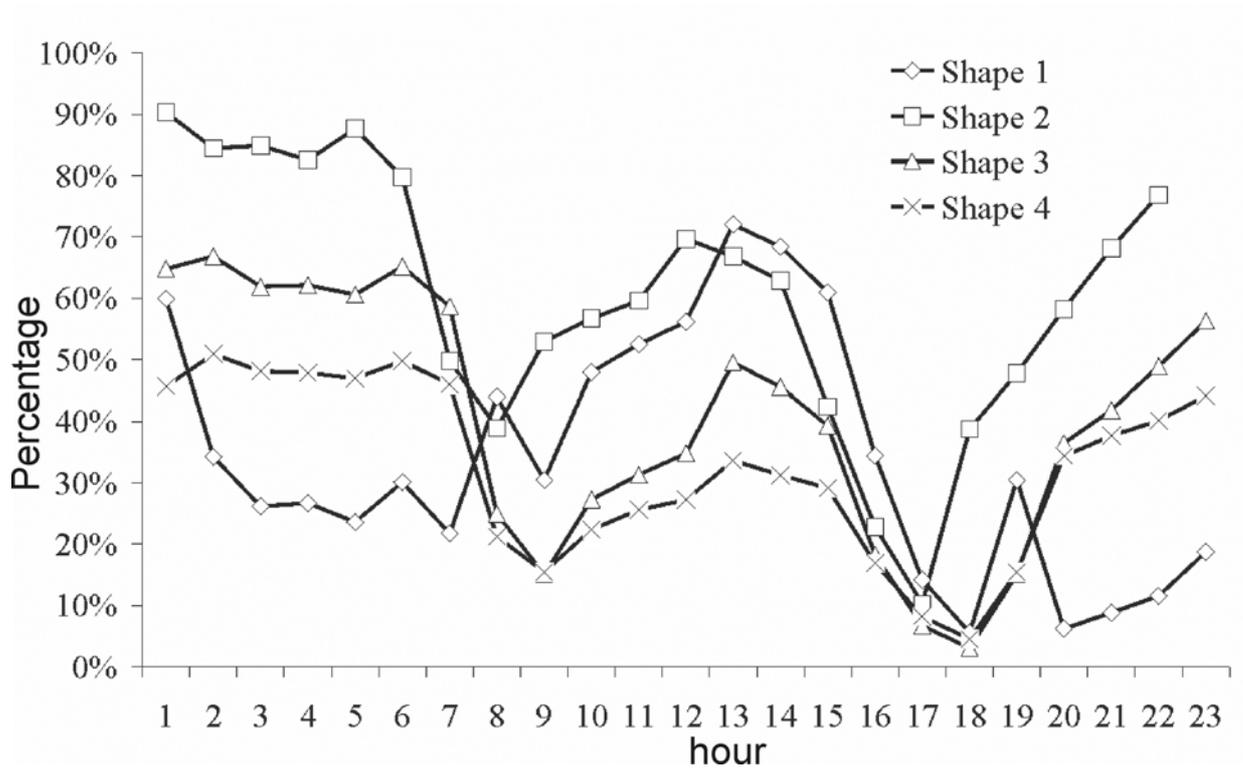

Figure 11: Hourly evolution of the percentage of well-ventilated volume in D5

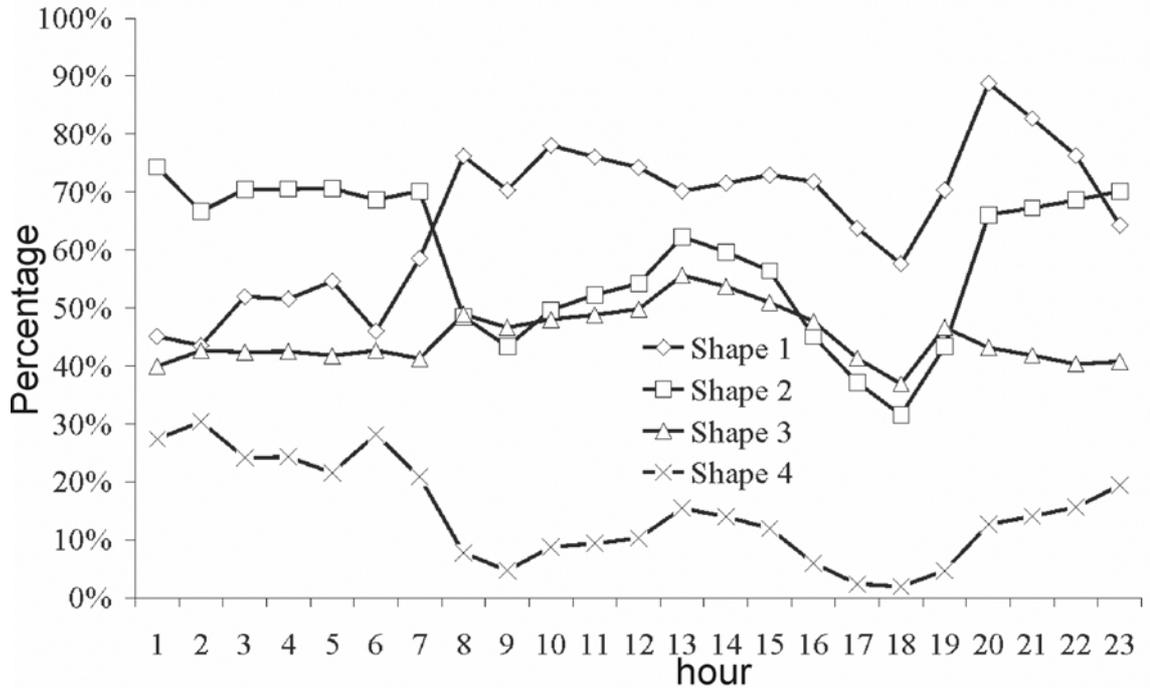

Figure 12: Hourly evolution of the percentage of well-ventilated volume in D8

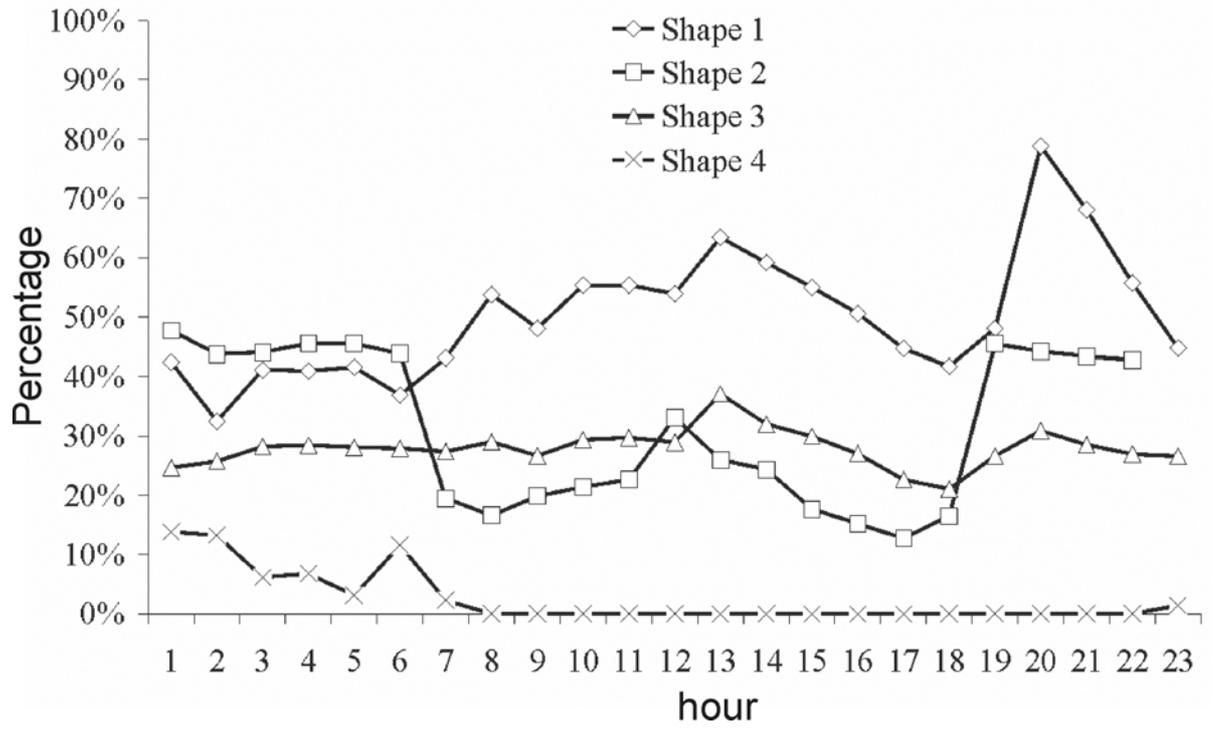

Figure 13: Hourly evolution of the percentage of well-ventilated volumes in D11

| Domain (D) | $X_{min}$ | $X_{max}$ | $Y_{min}$ | $Y_{max}$ | $Z_{min}$ | $Z_{max}$ | Plan H | Plan V |
|---|---|---|---|---|---|---|---|---|
| 1 | -1.5 | 1.5 | -1.5 | 1.5 | 0.0 | 2.8 | | |
| 2 | -1.2 | 1.2 | -1.2 | 1.2 | 0.0 | 2.0 | | |
| 3 | -1.2 | 1.2 | -1.2 | 1.2 | 0.5 | 1.0 | | |
| 4 | -1.2 | 1.2 | -1.2 | 1.2 | 1.0 | 1.5 | | |
| 5 | -1.2 | 1.2 | -1.2 | 1.2 | 0.3 | 0.6 | | |
| 6 | -1.2 | 1.2 | -1.2 | 1.2 | 0.6 | 1.8 | | |
| 7 | -1.2 | 1.2 | -1.2 | 1.2 | 1.2 | 1.5 | | |
| 8 | -1.2 | 1.2 | -1.2 | 1.2 | 1.1 | 1.8 | | |
| 9 | 0.0 | 1.2 | 0.0 | 1.2 | 0.7 | 1.5 | | |
| 10 | 0.0 | 1.2 | -1.2 | 0.0 | 0.7 | 1.5 | | |
| 11 | -1.2 | 0.0 | 0.0 | 1.2 | 0.7 | 1.5 | | |
| 12 | -1.2 | 0.0 | -1.2 | 0.0 | 0.7 | 1.5 | | |

Table 1: Evolution fields defined in the volume of the room

| Number | Shape | Description |
|---|---|---|
| 1 | 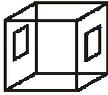 | Initial test building 3.2mx3.2x3.2m<br>2 large openings: 1.73mx1m |
| 2 | 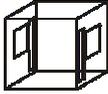 | 2 large openings: 1mx1m<br>2 vertical openings: 0.4mx1.8m |
| 3 | 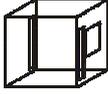 | 2 large openings: 1mx1m<br>1 vertical opening: 0.4mx1.8m |
| 4 | 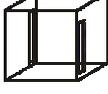 | 2 vertical openings: 0.4mx1.8m |

Table 2: Representation of the test building and the opening modifications